\documentclass{ws-rv9x6}
\usepackage{subfigure}   
\usepackage{ws-rv-thm}   
\usepackage{ws-rv-van}   
\makeindex
\usepackage{graphicx}
\usepackage{amssymb}
\usepackage{latexsym,graphicx,amssymb,amsmath,mathrsfs}
\usepackage{setspace,bm}
\usepackage{hyperref}

\newcommand{\diff}{\mathrm{d}}
\newcommand{\p}{\partial}
\newcommand{\ve}{\varepsilon}
\newcommand{\Diff}{{\mathcal{D}}}

\newcommand{\up}{\uparrow}
\newcommand{\down}{\downarrow}
\newcommand{\be}{\begin{equation}}      
\newcommand{\ee}{\end{equation}}      
\newcommand{\bea}{\begin{eqnarray}}      
\newcommand{\eea}{\end{eqnarray}}    
     
\newcommand{\rt}[1]{{}}

\newcommand{\fl}{}

\begin{document}

\chapter{Multi-Regulator Functional Renormalization Group \\ for Many-Fermion Systems}
\label{ra_ch1}

\author{Yuya Tanizaki$^{*}$ and Tetsuo Hatsuda$^{\dagger}$}

\address{$^*$ RIKEN BNL Research Center, Brookhaven National Laboratory,  Upton, NY 11973-5000, USA  \\
yuya.tanizaki@riken.jp}

\address{$^{\dagger}$ iTHES Research Group and Nishina Center, RIKEN, Wako 351-0198, Japan \\
thatsuda@riken.jp}

\begin{abstract}
We propose a method of multi-regulator functional renormalization group (MR-FRG) which is a novel formulation of functional renormalization group with multiple infrared regulators.
It is applied to a two-component fermionic system with an attractive contact interaction  to study crossover phenomena between the Bardeen-Cooper-Schrieffer (BCS) phase and the Bose-Einstein condensation (BEC) phase. 
To control both the fermionic one-particle excitations and the bosonic collective excitations,
  IR regulators are introduced, one for the fermionic two-point function and another for the four-fermion vertex.
  It is shown that  the Nozi\`eres-Schmitt-Rink (NSR) theory, which is successful to capture 
  qualitative features of the BCS-BEC crossover,  can be derived  from  MR-FRG.
 Some aspects of  MR-FRG to go beyond the NSR theory are also discussed. 
\end{abstract}
\body


\section{Introduction}\label{sec:intro}
Functional renormalization group (FRG) \cite{wetterich1993exact,Morris1,ellwanger1994flow} 
is a pragmatic realization of Wilson's idea of renormalization group \cite{Wilson}. 
Functional implementation of coarse-graining realizes an exact evolution of an effective action involving quantum and thermal fluctuations, and enables us to calculate observables of various quantum systems. 
Especially for interacting many fermions, the fermionic FRG  
provides a systematic way to analyze the system without \textit{a priori} knowledge on their ground state properties \cite{shankar1994renormalization,Salmhofer:2001tr,RevModPhys.84.299} 
  and is complementary to  bosonized FRG where auxiliary bosonic fields are introduced\cite{birse2005pairing,PhysRevB.78.174528, PhysRevA.81.063619}. 

We have previously  studied 
the Bardeen-Cooper-Schrieffer (BCS) to Bose-Einstein condensation (BEC) crossover of the two-component fermionic system with an attractive contact interaction \cite{PhysRev.186.456,leggett1980diatomic}
 by treating the BCS regime and the BEC regime separately on the basis of fermionic FRG:
In the BCS side,  the  Gorkov and Melik-Barkhudarov correction \cite{gorkov1961contribution} to the
BCS theory has been 
derived from fermionic FRG by  introducing an infrared (IR) regulator for the two-point function \cite{Tanizaki:2013doa,Tanizaki:2013fba},  while, in the BEC side,  fermionic FRG with an IR regulator for the four-fermion vertex has been  introduced   \cite{Tanizaki:2013yba}.

The purpose of this paper is to propose a new fermionic FRG with multiple IR regulators, which we call
 muti-regulator functional renormalization group (MR-FRG),   to control the 
low-energy fermionic and bosonic excitations simultaneously, so that it can treat 
the whole region of the BCS-BEC crossover. To show how our formalism works,
 we discuss its relation to the Nozi\`eres and Schmitt-Rink (NSR) theory  \cite{nozieres1985bose}, the
  conventional formalism to treat the BCS-BEC crossover with pairing fluctuations. 

This paper is organized as follows. 
In Sec.\ref{sec:formalism}, we consider a model for the two-component fermionic system, and introduce a new formulation of
 FRG with  two different  IR regulators and   two-parameter flow equations. 
In Sec.\ref{sec:formal}, formal properties of the flow equations are discussed. After carrying out  the 
ultraviolet (UV) renormalization of the flow equations,
we optimizate the RG flow by choosing appropriate  IR regulators  under the vertex expansion of the flow equations. 
In Sec.\ref{sec:nsr}, we derive the NSR theory for the BCS-BEC crossover based on MR-FRG. We first study the RG flow of the four-fermion vertex and observe that one-particle fermion excitations decouple from the flow equation. Using this property, we can solve the flow equation of the self-energy, whose solution correctly reproduces the number equation of the NSR theory. 
Sec.\ref{sec:summary} is devoted to summary and perspectives.

\section{Fermionic FRG with multiple IR regulators}\label{sec:formalism}
Let us consider the following action with non-relativistic two-component fermions $\Psi=\left(\begin{array}{c} \Psi_{\up}\\\Psi_{\down}\end{array}\right)$: 
\be
S[\overline{\Psi},\Psi]=\int_0^{\beta}\diff \tau\int \diff^3\bm{x}
\left[\overline{\Psi}\left(\p_{\tau}-{\nabla^2\over 2m}-\mu\right)\Psi(x)
+g\overline{\Psi}_{\up}\overline{\Psi}_{\down}\Psi_{\down}\Psi_{\up}(x)\right],
\label{form01}
\ee
with $\beta(=1/T)$ the inverse temperature, $\mu$ the chemical potential, $m$ the fermion mass, and $g$ the bare coupling constant of the contact interaction. The ultraviolet (UV) regularized form of (\ref{form01}) in the momentum space is given by 
\bea
S[\overline{\Psi},\Psi]&=&\int_p^{(T)} \overline{\Psi}_{p} G^{-1}(p) \Psi_{p}\nonumber\\
&+&g(\Lambda) \int_p^{(T)} e^{-ip^0 0^+}\int_{q,q'\le \Lambda}^{(T)}
\overline{\Psi}_{\up,{p\over 2}+q}\overline{\Psi}_{\down,{p\over 2}-q}
\Psi_{\down,{p\over 2}-q'}\Psi_{\up,{p\over 2}+q'},
\label{form02}
\eea
where $G^{-1}(p)=i p^0+\bm{p}^2/2m-\mu$ with $p=(p^0,\bm{p})$, $\psi_{\sigma,p}$ denotes the Fourier expansion
coefficient of $\psi_{\sigma}(x)$, $\int_p^{(T)}=\int{\diff^3\bm{p}\over (2\pi)^3}{1\over \beta}\sum_{p^0}$, and $\Lambda$
 denotes a  UV cutoff for  the spatial momenta, $\bm{q}$ and $\bm{q}'$. 
 
The fermionic field theory (\ref{form02}) contains UV-divergence only in the loop of particle-particle scattering, 
\be
\Pi(p)=\int_{l\le \Lambda}^{(T)}{1\over G^{-1}({p\over 2}+l)G^{-1}({p\over 2}-l)}={m\Lambda\over 2\pi^2}+\mathcal{O}(1).  
\label{vac02}
\ee
Such a divergence can be absorbed into the bare coupling $g(\Lambda)$ and renormalized into  
 the physical scattering length $a_s$;
\bea
\frac{m}{4\pi a_s} = g^{-1}(\Lambda) + \frac{m \Lambda}{2\pi^2}.
\label{eq:RNO}
\eea
Observables are
 written in terms of $m$, $a_s$, $T$ and $\mu$ by absorbing all the  UV divergences through Eq.(\ref{eq:RNO}) 
 and taking $\Lambda \rightarrow \infty$. 
  
In order to describe the BCS-BEC crossover by fermionic FRG,
we need to control both fermionic one-particle excitations and bosonic collective excitations. 
Since fermionic excitations are described by the two-point Green function, we introduce an additional action, 
\be
\delta S^{(f)}_{k_1}[\overline{\Psi},\Psi]=\int^{(T)}_p \overline{\Psi}_p R^{(f)}_{k_1}(\bm{p}) \Psi_p, 
\label{form03}
\ee
where $R^{(f)}_{k_1}$ regulates fermionic one-particle excitations with energy smaller than $k_1^2/2m$. 
On the other hand, the collective bosonic excitation is described by the pole of the four-point Green function. In order to suppress bosonic excitations with  energy less than $k_2^2/4m$, we introduce the following action \cite{Tanizaki:2013yba},
\be
\delta S^{(b)}_{k_2}[\overline{\Psi},\Psi]=\int^{(T)}_p {g^2 R^{(b)}_{k_2}(\bm{p}) e^{-ip^0 0^+}\over 1-g R^{(b)}_{k_2}(\bm{p})}\int^{(T)}_{q,q'\le \Lambda} \overline{\Psi}_{\up,{p\over 2}+q}\overline{\Psi}_{\down,{p\over 2}-q}\Psi_{\down,{p\over 2}-q'}\Psi_{\up,{p\over 2}+q'}, 
\label{form04}
\ee
with $R^{(b)}_{k_2}$ being an vertex  IR regulator. For later convenience, we define
\be
g_{k_2}(p) \equiv \frac{g^2 R^{(b)}_{k_2}(\bm{p})}{1-g R^{(b)}_{k_2}(\bm{p})}.
\ee 
Note that adding this vertex IR regulator introduces an effective shift of the inverse coupling,
  $g^{-1} \rightarrow g^{-1} -R_{k_2}^{(b)}(\bm{p})$.
   
With an abbreviated notation $\phi\equiv (\overline{\Psi},\Psi)$, 
 the Schwinger functional $W_{k_1,k_2}[J]$ depending on two scales  $(k_1,k_2)$ can be defined as
\be
W_{k_1,k_2}[J]= \ln \ \int \Diff \phi \exp\left[-\left(S[\phi]+\delta S^{(f)}_{k_1}[\phi]+\delta S^{(b)}_{k_2}[\phi]\right)+J^{\alpha} \phi_{\alpha}\right]. 
\label{form05}
\ee
In this expression, $\alpha$ runs over all the arguments of  $\phi$ (spacetime coordinates $x$, internal degrees of freedom $\sigma$, and the global U$(1)$ charges $\pm 1$ for $\overline{\Psi}$ and $\Psi$). 
The $(k_1,k_2)$-dependent one-particle-irreducible (1PI) effective action $\Gamma_{k_1,k_2}[\varphi]$ 
 with $\varphi \equiv (\overline{\psi},\psi)$
 is defined as the Legendre transform of the Schwinger functional, 
\be
\Gamma_{k_1,k_2}[\varphi]+\delta S^{(f)}_{k_1}[\varphi]= J^{\alpha}[\varphi] \varphi_{\alpha}-W_{k_1,k_2}\left[J[\varphi]\right], 
\label{form06}
\ee
where $J[\varphi]$ is the solution of $\delta_L W_{k_1,k_2}[J]/\delta J=\varphi$ with  $\delta_L/\delta J$ being the left functional derivative. This is the generating functional of the 1PI effective vertices at the scale $(k_1,k_2)$. 

In terms of the parameter $k_1$, the 1PI effective action $\Gamma_{k_1,k_2}[\varphi]$ obeys the flow equation\cite{wetterich1993exact,Morris1,ellwanger1994flow},
\be
\p_{k_1}\Gamma_{k_1,k_2}[\varphi]={1\over 2}\mathrm{STr}\left[{\p_{k_1} R^{(f)}_{k_1}\over \Gamma^{(2)}_{k_1,k_2}[\varphi]+R^{(f)}_{k_1}}\right], 
\label{form07}
\ee
where $\Gamma^{(2)}_{k_1,k_2}[\varphi]={\delta_L\over \delta \varphi}{\delta_R\over \delta \varphi}\Gamma_{k_1,k_2}[\varphi]$ is the field-dependent propagator, and $\mathrm{STr}$ denotes the supertrace in the functional space.
For  fixed  and large UV cutoff $\Lambda$, the initial condition of the flow equation is given by
 $\Gamma_{k_1=\Lambda \to \infty,k_2}[\varphi]=S[\varphi]+\delta S^{(b)}_{k_2}[\varphi]$ 
 with a possible field-independent part. 
For this boundary condition to be satisfied, we need to choose $R^{(f)}_{k_1=\Lambda \to \infty}=\infty$ and
 $R^{(f)}_{k_1=0}=0$.

\begin{figure}[tb]
\bea
\p_{k_1} \Gamma_{k_1,k_2}[\varphi]&=&\parbox{3.5em}{\includegraphics[width=3.5em]{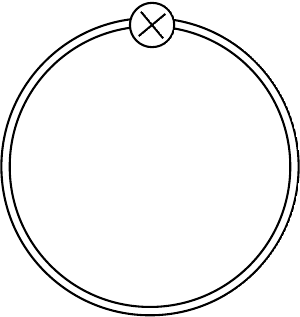}} \nonumber \\
\vspace{0.2cm}
\p_{k_2} \Gamma_{k_1,k_2}[\varphi]&=&
\parbox{2.6em}{\includegraphics[width=2.6em]{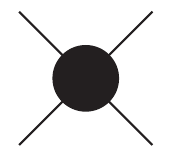}}
+\parbox{3.5em}{\includegraphics[width=3.5em]{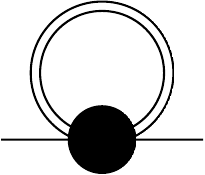}}
+\parbox{7em}{\includegraphics[width=7em]{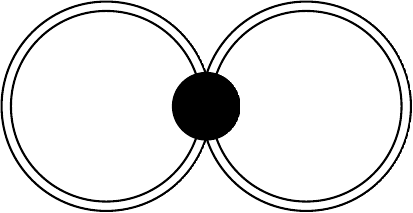}} \nonumber \\
& &+\parbox{7em}{\includegraphics[width=7em]{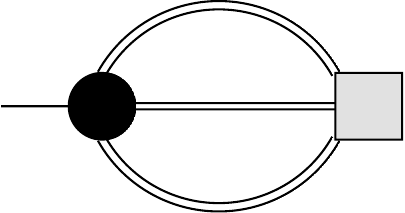}} 
 + \parbox{6em}{\includegraphics[width=6em]{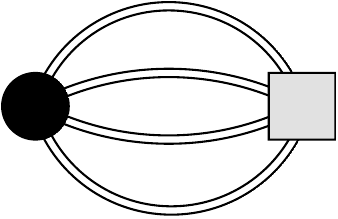}}
+\parbox{6em}{\includegraphics[width=6em]{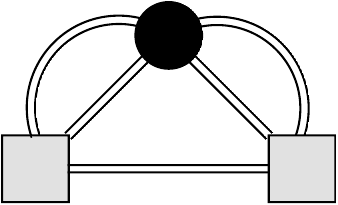}}\nonumber
\eea
\caption{The flow equations for the effective action, $\Gamma_{k_1,k_2}[\varphi]$, in terms of $k_1$ and $k_2$. 
The thin lines represent the field $\varphi$,  the double lines represent the field-dependent propagator,  $G_{k_1,k_2} [\varphi]$,  
the circle with cross represents $\p_{k_1} R^{(f)}_{k_1}$, 
the filled circles  represent $- \p_{k_2} g_{k_2}$, and the square boxes represent the field-dependent 1PI effective vertice,  
$\Gamma_{k_1,k_2}^{(3,4)}[\varphi]$. }
\label{fig:flow_eff1}
\end{figure}

The $k_2$-dependence of $\Gamma_{k_1,k_2}$ is obtained from the following flow equation~\cite{Tanizaki:2013yba},
\bea\fl
\p_{k_2} \Gamma_{k_1,k_2}[\varphi]&=&{1\over 4!}\p_{k_2} g_{k_2}^{\alpha_1\alpha_2\alpha_3\alpha_4}
\Big(\varphi_{\alpha_1}\varphi_{\alpha_2}\varphi_{\alpha_3}\varphi_{\alpha_4}\nonumber\\
&+&6\varphi_{\alpha_1}\varphi_{\alpha_2}G_{k_1 k_2,\alpha_3\alpha_4}+3G_{k_1 k_2,\alpha_1\alpha_2}G_{k_1 k_2,\alpha_3\alpha_4}\nonumber\\
&+&4\varphi_{\alpha_1}G_{k_1 k_2,\alpha_2\beta_2}G_{k,\alpha_3\beta_3}G_{k_1 k_2,\alpha_4\beta_4}\Gamma_{k_1,k_2}^{(3),\beta_2\beta_3\beta_4}\nonumber\\
\fl&+&G_{k_1 k_2,\alpha_1\beta_1}G_{k_1 k_2,\alpha_2\beta_2}G_{k_1 k_2,\alpha_3\beta_3}G_{k_1 k_2,\alpha_4\beta_4}\Gamma_{k_1,k_2}^{(4),\beta_1\beta_2\beta_3\beta_4}\nonumber\\
\fl&+&3G_{k_1 k_2,\alpha_1\beta_1}G_{k_1 k_2,\alpha_2\beta_2}G_{k_1 k_2,\alpha_3\beta_3}G_{k_1 k_2,\alpha_4\beta_4}G_{k_1 k_2,\gamma_1\gamma_2}\nonumber\\
&&\times\Gamma_{k_1,k_2}^{(3),\beta_1\beta_2\gamma_1}\Gamma_{k_1,k_2}^{(3),\gamma_2\beta_3\beta_4}\Big),
\label{form08}
\eea
with $G_{k_1,k_2}[\varphi]=\left(\Gamma^{(2)}_{k_1,k_2}[\varphi]+R^{(f)}_{k_1}\right)^{-1}$ being the field dependent propagator, and $\Gamma_{k_1,k_2}^{(n)}[\varphi]$ being the $n$-th functional derivative of $\Gamma_{k_1,k_2}[\varphi]$. 
The boundary condition of the RG flow is specified by requiring $R^{(b)}_{k_2=\Lambda \to\infty}\to \infty$ and $R^{(b)}_{k_2=0}=0$.  For large  UV cutoff $\Lambda$, the 1PI effective action $\Gamma_{k_1,k_2}[\varphi]$ converges to the free action in the limit $k_2=\Lambda \to \infty$, because $g + g_{k_2=\Lambda \to \infty}= g-g =0$.

Diagrammatic expressions of the flow equations in MR-FRG,  (\ref{form07}) and  (\ref{form08}),  are
 shown in Fig.\ref{fig:flow_eff1}, where  the double lines represent the field-dependent propagator,  $G_{k_1,k_2} [\varphi]$.

\section{Formal aspects of the flow equations in MR-FRG}\label{sec:formal}
Eqs. (\ref{form07}) and (\ref{form08}) constitute the basic equations of MR-FRG. 
In this section, we study formal aspects of the  flow equations in MR-FRG, and consider optimization of the flow with respect to the IR regulators. 
\subsection{UV renormlization  of the flow equations}\label{subsec:uv_indep}

\begin{figure}[t]
\centering
$\parbox{9em}{\includegraphics[width=9em]{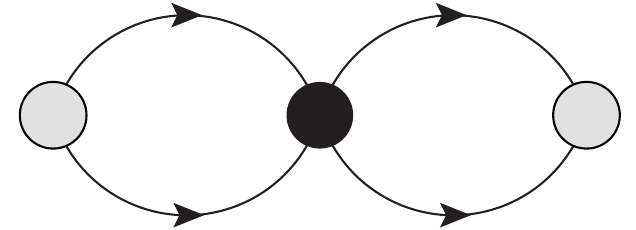}}=-\p_{k_2}R^{(b)}_{k_2}$
\caption{The UV-finite diagram which appears in the flow equation.  The grey circles denote the bare  vertex $g$, while the filled  circle represents $-\p_{k_2}g_{k_2}$. The internal lines with arrows are the field-independent propagator, $G_{k_1,k_2}[\varphi=0]$.}
\label{fig:subdiagram}
\end{figure}

In Eq.(\ref{form07}), the derivative of the fermion IR regulator $\p_{k_1}R^{(f)}_{k_1}$ introduces 
a UV-cutoff at the energy scale $k_1^2/2m$ in the loop integration, and thus it is free from UV divergences in the 
limit, $\Lambda \rightarrow \infty$.  Let us now show that Eq.(\ref{form08}) does not have UV divergences too.
 
\begin{figure}[t]
\bea
\p_{k_2} \Gamma_{k_1,k_2}[\varphi]&=&
\parbox{6em}{\includegraphics[width=6em]{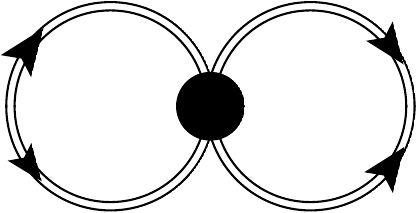}}
+\parbox{5em}{\includegraphics[width=5em]{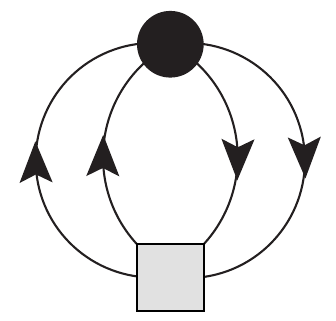}}\nonumber\\
& & +\parbox{5em}{\includegraphics[width=5em]{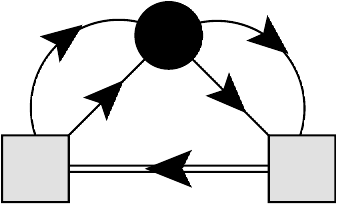}}
+\parbox{5em}{\includegraphics[width=5em]{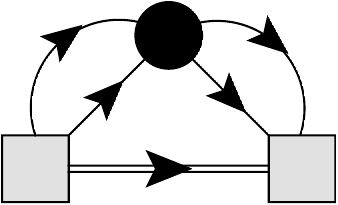}}
+\parbox{5em}{\includegraphics[width=5em]{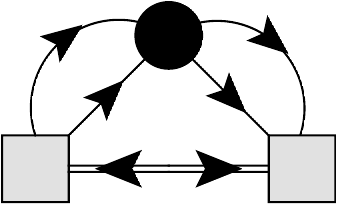}}
+\parbox{5em}{\includegraphics[width=5em]{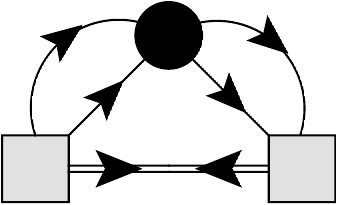}}\nonumber
\eea
\caption{UV-renormalized $k_2$-flow equation of the effective action. The field-independent propagator ($G_{k_1,k_2}[0]$)
 and the field-independent propagator ($G_{k_1,k_2}[\varphi]$) are represented by the single line and the double  line, respectively.}
\label{fig:flow_eff2}
\end{figure}

 We first note that the bare coupling appears in the flow equation as $\p_{k_2} g_{k_2}=g^2 \p_{k_2} R^{(b)}_{k_2}/(1-g R^{(b)}_{k_2})^2$. Therefore, together with the renomalization condition, Eq.(\ref{eq:RNO}),  we have 
\bea
 \p_{k_2}g_{k_2} \xrightarrow[\Lambda\to \infty]{ } \left( \frac{2\pi^2}{m\Lambda} \right)^2 \p_{k_2}R^{(b)}_{k_2}
  \sim O(1/\Lambda^2).
\label{eq:pkg}
\eea
Now, the UV-finite sub-diagrams originate only from particle-particle scattering contribution 
(see Fig.\ref{fig:subdiagram})\footnote{From now on, we will  attach arrows to fermion lines  in order to identify  whether 
 the fermion is emitted or absorbed at each vertex. }
  in the present  non-relativistic system,
\be
\lim_{\Lambda\to \infty}[\Pi_{k_1,k_2}(p)]^2 \left[-\p_{k_2}g_{k_2}(p)\right]=-\p_{k_2}R^{(b)}_{k_2}(\bm{p}),
\ee
where $\Pi_{k_1,k_2}(p)$ is a particle-particle 1-loop amplitude of  the field-independent propagator, 
$G_{k_1,k_2}[0]=\left(\Gamma^{(2)}_{k_1,k_2}[0]+R^{(f)}_{k_1}\right)^{-1}$.
Since all the other sub-diagrams are suppressed by the powers of $1/\Lambda$,
the $k_2$-flow equation in Fig.~\ref{fig:flow_eff1} can be much simplified in the UV-limit  as shown in Fig.\ref{fig:flow_eff2}.  
To make clear distinction between the field-independent propagator $G_{k_1,k_2}[0]$
and the  field-dependent propagator $G_{k_1,k_2}[\varphi]$,
we represent the former (latter) by a single (double)  line in this figure.

\subsection{Vertex expansion of the flow equations}\label{subsec:vertex}
\begin{figure}[t]
\bea
\partial_{k_1} \parbox{5em}{\includegraphics[width=5em]{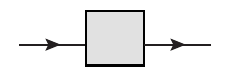}}
&=&\widetilde{\partial}_{k_1} \parbox{5em}{\includegraphics[width=5em]{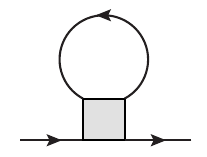}} \nonumber \\
\displaystyle \partial_{k_1} \parbox{5em}{\includegraphics[width=5em]{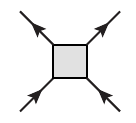}}
&=&\widetilde{\partial}_{k_1}\Biggl(
\parbox{3em}{\includegraphics[width=3em]{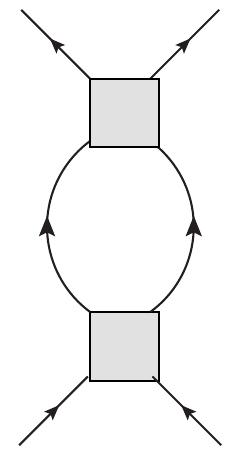}}+
\parbox{5em}{\includegraphics[width=5em]{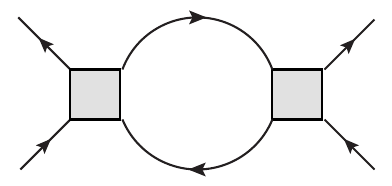}}
\Biggr) \nonumber
\eea
\caption{The flow equations for the self energy $\Sigma_{k_1,k_2}$ and the four-point vertex $\Gamma_{k_1,k_2}^{(4)}$ in terms of $k_1$. 
\label{fig:flow_4pt}}
\end{figure}

In order to solve the flow equations (\ref{form07}) and (\ref{form08})  for each 1PI vertex function, we need to perform the vertex expansion of $\Gamma_{k_1,k_2}[\overline{\psi},\psi]$. For simplicity, we assume the symmetry-unbroken phase in the following, so that each vertex couples to the same number of  $\psi$ and  $\overline{\psi}$. 
Then we consider the following vertex expansion of $\Gamma_{k_1,k_2}$ up to the fourth order,
\bea\fl
\Gamma_{k_1,k_2}[\overline{\psi},\psi]
&=&\int_p^{(T)}\overline{\psi}_p [G^{-1}-\Sigma_{k_1,k_2}](p) \psi_p\nonumber\\
&+&\int_{p,q,q'}^{(T)}\Gamma_{k_1,k_2}^{(4)}(p,q,q')\overline{\psi}_{\up,{p\over2}+q}\overline{\psi}_{\down,{p\over2}-q} \psi_{\down,{p\over2}-q'}\psi_{\up,{p\over2}+q'},
\label{ex01}
\eea
where  $\Sigma_{k_1,k_2}$ and  $\Gamma_{k_1,k_2}^{(4)}$ are the self-energy and the four-point vertex, respectively. 
The flow equations in terms of $k_1$ with this truncation are shown in Fig.\ref{fig:flow_4pt} with $\widetilde{\p}_{k_1}$ the $k_1$-derivative acting only on the explicit $k_1$-dependence of the fermionic IR regulator $R^{(f)}_{k_1}$.  Their analytic expressions
are  obtained by substituting (\ref{ex01}) into  (\ref{form07}):
\be
\label{flow_se}
\p_{k_1}\Sigma_{k_1,k_2}(p)=\widetilde{\p}_{k_1} \int_l^{(T)} e^{-i l^0 0^+}
{\Gamma_{k_1,k_2}^{(4)}\left(p+l\right)\over [G^{-1}-\Sigma_{k_1,k_2}+R^{(f)}_{k_1}](l)} ,
\ee
\bea
&&-\p_{k_1} \Gamma_{k_1,k_2}^{(4)}(p)=\widetilde{\p}_{k_1} \Bigg[\int_l^{(T)}{\Gamma_{k_1,k_2}^{(4)}(p)\Gamma_{k_1,k_2}^{(4)}(p) \over \prod_{\pm}\left([G^{-1}-\Sigma_{k_1,k_2}+R^{(f)}_{k_1}]\left({p\over 2}\pm l\right)\right)}
\nonumber\\
\fl
&&+{1\over 2}\sum_{\pm}\int_l^{(T)}
\scalebox{0.9}{$\displaystyle
{\Gamma_{k_1,k_2}^{(4)}\left({p\over 2}+q+l\right) \Gamma_{k_1,k_2}^{(4)}\left({p\over 2}\pm q'+l\right)\over  [G^{-1}-\Sigma_{k_1,k_2}+R^{(f)}_{k_1}](l)[G^{-1}-\Sigma_{k_1,k_2}+R^{(f)}_{k_1}](q\pm q'+l)}
$}
\Bigg]. 
\label{flow_4pt}
\eea
Here,  relative momenta  dependence of the 1PI four-point vertex is not shown explicitly just for simplicity. 

\begin{figure}[t]
\bea
\p_{k_2} \parbox{5em}{\includegraphics[width=5em]{self_energy_wo_momenta}}
&=&\parbox{9em}{\includegraphics[width=9em]{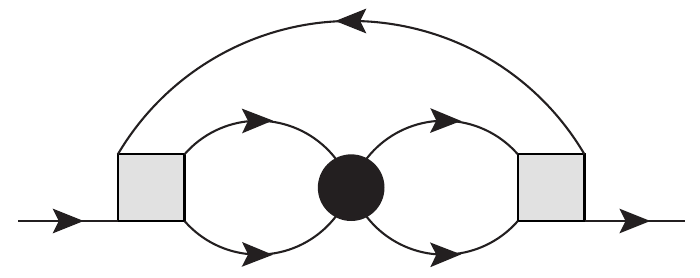}} \nonumber \\
\p_{k_2}\parbox{4em}{\includegraphics[width=4em]{4pt_wo_momenta}}
&=& \parbox{9em}{\includegraphics[width=9em]{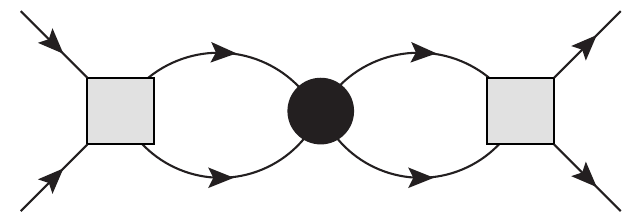}}
+\parbox{9em}{\includegraphics[width=9em]{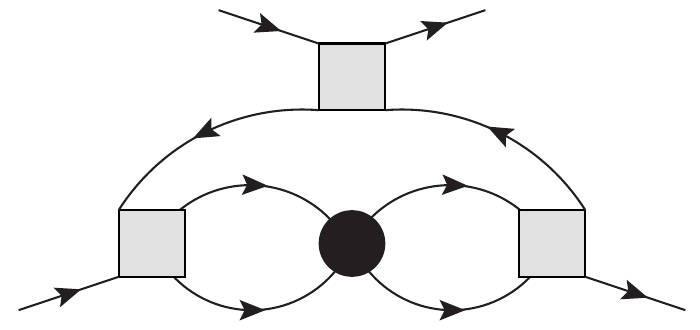}} \nonumber
\eea
\caption{Flow equation of the self energy $\Sigma_{k_1,k_2}$ and the four-point vertex $\Gamma_{k_1,k_2}^{(4)}$ in terms of $k_2$.}
\label{fig:vflow_4pt}
\end{figure}

The flow equations 
for  $\Sigma_{k_1,k_2}$ and $\Gamma^{(4)}_{k_1,k_2}$ in terms of $k_2$ can be obtained
by substituting the vertex expansion (\ref{ex01}) into the flow equation in Fig.\ref{fig:flow_eff2}.
 Their diagrammatic expressions are shown in Fig.\ref{fig:vflow_4pt}, and the analytic forms are given by 
\be
\label{vflow_se}
\p_{k_2}\Sigma_{k_1,k_2}(p)=\int_l^{(T)} e^{-i l^0 0^+}
{\left(\Gamma_{k_1,k_2}^{(4)}\left(p+l\right)\right)^2 \p_{k_2}R^{(b)}_{k_2}(\bm{p}+\bm{l})\over [G^{-1}-\Sigma_{k_1,k_2}+R^{(f)}_{k_1}](l)} ,
\ee
\bea
&&-\p_{k_2} \Gamma_{k_1,k_2}^{(4)}(p)
=-\left(\Gamma^{(4)}_{k_1,k_2}(p)\right)^2 \p_{k_2}R^{(b)}_{k_2}(\bm{p})
\nonumber\\
\fl
&&+\sum_{\pm}\int_l^{(T)}
\scalebox{0.9}{$\displaystyle
{\Gamma_{k_1,k_2}^{(4)}\left({p\over 2}+q+l\right) \left(\Gamma_{k_1,k_2}^{(4)}\left({p\over 2}\pm q'+l\right)\right)^2 \p_{k_2}R^{(b)}_{k_2}({\bm{p}\over 2}\pm \bm{q}'+\bm{l})\over  [G^{-1}-\Sigma_{k_1,k_2}+R^{(f)}_{k_1}](l)[G^{-1}-\Sigma_{k_1,k_2}+R^{(f)}_{k_1}](q\pm q'+l)}
$}. 
\label{vflow_4pt}
\eea
When we write down the full momentum dependence of this flow equation, we should notice that only the relative-momentum independent part of each 1PI vertex can couple to $\p_{k_2} g_{k_2}$ in the limit $\Lambda \to \infty$. Such momentum independent parts can be extracted by putting the corresponding relative momentum to be infinity. 

After solving the flow equations (\ref{flow_se}-\ref{vflow_4pt}), we apply the Thouless criterion for the superfluid phase transition \cite{thouless1960perturbation}, 
\be
{1\over \Gamma^{(4)}_{0,0}(p=0)}=0, 
\label{eq:critical}
\ee
which corresponds to the gap equation at $T=T_c$.  On the other hand, the number density $n$ is obtained by 
\be
n=-2\int^{(T)}_p {e^{-i p^0 0^+}\over G^{-1}(p)-\Sigma_{0,0}(p)}. 
\label{eq:number}
\ee
By combining  (\ref{eq:critical}) and (\ref{eq:number}), 
 one  eventually  obtains   
 $T_c/\ve_F$ as a function of $(k_F a_s)^{-1}$ with  $k_F=(3\pi n)^{1/3}$ and $\ve_F=k_F^2/2m$ . 

\subsection{Optimized flow and IR regulators}\label{subsec:regulator}
There  are many  degrees of freedom in the choice of IR regulators. The idea of optimizing
 the FRG flow  is to use such degrees of freedom to construct a systematic approximation scheme. 
The optimized IR regulator depends naturally  on the approximation scheme and the order of truncation. 
For bosonic theories, regulator dependence of the approximate FRG flow has been studied in details within the local potential approximation (LPA) \cite{litim2000optimisation,litim2001optimized,Pawlowski}. 
According to these studies, the optimized IR regulators have been found to share 
 the following properties besides the boundary condition at $k=\infty$: 
\begin{enumerate}
\item[(i)] IR regulator $R_k$ makes the system gapped by a typical energy scale $k^2/2m$ with $k$ being the scale parameter. 
\item[(ii)] High-energy excitations with the energy larger than $k^2/2m$ decouple from the flow equation. 
\end{enumerate}
With these properties, low-energy excitations are equally gapped and high-energy excitations do not
affect the  low-energy dynamics after renormalization. This helps us to establish a series of systematic
 truncations of the flow equations. 

For the system treated in this paper, 
there are two different degrees of freedom to be gapped. The first one is the one-particle fermion excitation. 
The above two condition for such excitations can be  satisfied 
by taking the Litim-type optimized regulator,\cite{litim2000optimisation,litim2001optimized}
\begin{equation}
R^{(f)}_{k_1}(\bm{p}) \equiv
Z_f\Big[{k_1^2\over 2m}\mathrm{sgn}(\xi(\bm{p}))-\xi(\bm{p})\Big]\theta\Big({k_1^2\over 2m}-|\xi(\bm{p})|\Big),
\label{eq:fermion_reg}
\end{equation}
where $\xi(\bm{p})$ denotes the excitation energy relative to the Fermi level, with $Z_f$ the wave-function renormalization of fermionic fields.  For the simplest case, $Z_f=1$ and $\xi(\bm{p})=\frac{\bm{p}^2}{2m} -\mu $, it is easy to check that this regulator opens an energy  gap by the amount
 $\pm \frac{k_1^2}{2m}$ near the Fermi surface, $\left |\frac{\bm{p}^2}{2m} -\mu \right|  \le   \frac{k_1^2}{2m}$.

The second one is the bosonic collective excitation given by the pole of $\Gamma_{k_1,k_2}^{(4)}(p)$.
According to the flow equation in Fig.\ref{fig:vflow_4pt}, $\p_{k_2}R^{(b)}_{k_2}$ can make high-energy collective excitations decouple from the flow equation. Therefore, the Litim-type regulator is again expected to satisfy the above two conditions,
\cite{litim2000optimisation,litim2001optimized}
\be
R^{(b)}_{k_2}\equiv Z_b\left({k_2^2\over 4m}-{\bm{p}^2\over 4m}\right) \theta \Big({k_2^2\over 4m}-{\bm{p}^2\over 4m}\Big), 
\label{eq:boson_reg}
\ee
where $Z_b$ is  the wave-function renormalization of the effective boson propagator. 
 For the simplest case, $Z_b=1$, this introduces a gap by $\frac{k_2^2}{4m}$ 
 at low momentum, $\frac{\bm{p}^2}{4m}   \le   \frac{k_2^2}{4m}$.

We can convert the two-parameter flow equations (\ref{flow_se}-\ref{vflow_4pt}) into a single-parameter flow equation by relating $k_1$ and $k_2$. 
In this paper, we simply take $k \equiv k_1=k_2 $, and adopt a simplified notation
 $\Gamma_k[\overline{\psi},\psi]\equiv \Gamma_{k,k}[\overline{\psi},\psi]$.  
 One may alternatively choose $k_1=c k_2$ with a constant  $c$ which can be determined e.g. by 
 applying the principle of minimal sensitivity \cite{stevenson1981optimized} to  some observable such as $T_c/\ve_F$. 
 
\section{NSR theory from MR-FRG}\label{sec:nsr}

In this section, we derive the NSR theory \cite{nozieres1985bose} on the basis of MR-FRG.
In the NSR theory, the particle-particle random phase approximation (pp-RPA) is adopted  to take into account  the 
effect of collective fermion pairs  in the free energy.  Then the critical temperature in the BCS-BEC crossover can be calculated as a function of $(k_F a_s)^{-1}$ by solving the following two equations,
\bea\fl \hspace{3em}
{1\over a_s}&=&-{2\over \pi}\int_0^{\infty}\sqrt{2m\ve}\diff\ve\left[{\tanh{\beta\over 2}(\ve-\mu)\over 2(\ve-\mu)}-{1\over 2\ve}\right], 
\label{eq:nsr_gap}\\\fl
n&=&-2\int^{(T)}_p e^{- ip^0 0^+}G(p)\nonumber\\
&-&{\p\over \p \mu}\int^{(T)}_p e^{-i p^0 0^+}\ln\left[1+{4\pi a_s\over m}\left(\Pi(p)-{m\Lambda\over 2\pi^2}\right)\right]. 
\label{eq:nsr_num}
\eea

\subsection{Pairing approximation on the four-point vertex}\label{subsec:vertex_flow}
We first consider the flow equation of the four-point 1PI vertex function, and identify
 an approximation corresponding to pp-RPA. 
Flow equations of the four-point 1PI vertex $\Gamma^{(4)}_{k_1,k_2}(p;q,q')$ are given in Figs.\ref{fig:flow_4pt} and \ref{fig:vflow_4pt}. As we have announced at the end of the previous section,
we  set  $k=k_1=k_2$ and analyze the flow equation for $\Gamma^{(4)}_{k}=\Gamma^{(4)}_{k,k}$.
 Let us consider only the particle-particle correlation in the flow equation of the four-point vertex by
  neglecting the self-energy correction: 
\bea\fl
-\p_{k} \Gamma_{k}^{(4)}(p)
&=&\int_l^{(T)}{-2\left(\Gamma_{k}^{(4)}(p)\right)^2\p_k R^{(f)}_{k}({\bm{p}\over 2}-\bm{l}) \over [G^{-1}+R^{(f)}_{k}]\left({p\over 2}+l\right)[G^{-1}+R^{(f)}_{k}]^2\left({p\over 2}-l\right)}\nonumber\\
&+&\left(\Gamma^{(4)}_{k}(p)\right)^2 \p_{k}R^{(b)}_{k}(\bm{p}).
\label{flow_vertex_nsr}
\eea
Under this approximation, the solution of the flow equation becomes the four-point vertex in the pp-RPA \cite{Tanizaki:2013doa,Tanizaki:2013yba}. 
Since the fermion dispersion relation is given by the bare propagator, parameters of the IR regulator $R^{(f)}_k$ in (\ref{eq:fermion_reg}) can be specified to be $Z_f=1$ and $\xi(\bm{p})=\bm{p}^2/2m-\mu$.  Therefore,
the analytic expression for the four-point vertex becomes 
\bea\fl\hspace{2em}
&&{1\over \Gamma_{k}^{(4)}(p)}
={m\over 4\pi a_s}-R^{(b)}_{k}(\bm{p})\nonumber\\
\fl&&+\int{\diff^3\bm{\ell}\over (2\pi)^3}\left[{1-\sum_{\pm}n_F\left({(\bm{p}/2\pm \bm{\ell})^2\over 2m}-\mu+R^{(f)}_{k}({\bm{p}\over 2}\pm \bm{\ell})\right)\over {\bm{\ell}^2\over m} +(i p^0+{\bm{p}^2\over 4m} -2\mu) +\sum_{\pm}R^{(f)}_{k}({\bm{p}\over 2}\pm \bm{\ell})}-{1\over {\bm{\ell}^2\over m}}\right]. 
\label{eq:4pt_rpa}
\eea

  Then the  Thouless criterion (\ref{eq:critical}) leads to 
\bea
0&=& \frac{1}{\Gamma_{0}^{(4)}(p=0)} \nonumber\\
&=&{m\over 4\pi a_s}+ \int{\diff^3\bm{\ell}\over (2\pi)^3}\left[{1-2n_F\left({\bm{\ell}^2\over 2m}-\mu\right)\over \bm{\ell}^2/m -2\mu} -{1\over \bm{\ell}^2/m}\right], 
\label{eq:thouless}
\eea
with $n_F$ the Fermi distribution function. 
Eq. (\ref{eq:thouless}) is identical to  Eq.(\ref{eq:nsr_gap}) in NSR theory. From (\ref{eq:thouless}), the chemical potential $\mu$ must be positive in the BCS region, $(k_F a_s)^{-1}\lesssim -1$, and negative in the BEC region, $(k_F a_s)^{-1}\gtrsim 1$. 
There exists low-energy fermions in the BCS side, while all the fermions are gapped due to the binding energy $1/2m a_s^2$ in the BEC side. 

Now, the non-trivial part in the derivation of the NSR theory comes from the pairing fluctuation in the number equation (\ref{eq:nsr_num}). 
In Sec.\ref{subsec:structure_k},  we first study the  $k$-dependence of the four-point 1PI vertex function.
Then  we derive the number equation  by analyzing the structure of the flow equation for the self-energy in  Sec.\ref{subsec:se_flow}.

\subsection{Structure of the $k$-dependent four-point 1PI vertex}\label{subsec:structure_k}

Since the following analysis depends on the sign of the chemical potential, we consider the BCS side  and BEC side
 separately. 
 
In the BCS side where the chemical potential $\mu$ is positive, 
the condition $\xi(\bm{q})=0$ defines a Fermi sphere $\{|\bm{q}|=\sqrt{2m\mu}\}$. Using the Thouless criterion (\ref{eq:thouless}), the equation (\ref{eq:4pt_rpa}) can be rewritten as 
\bea\fl
{-1\over \Gamma^{(4)}_k(p)}&=&-\int{\diff^3\bm{\ell}\over (2\pi )^3} \left[{1-\sum_{\pm}n_F\left(\xi({\bm{p}\over 2}\pm \bm{\ell})+R^{(f)}_{k}({\bm{p}\over 2}\pm \bm{\ell})\right)\over 2\xi(\bm{\ell}) +(i p^0+{\bm{p}^2\over 4m}) +\sum_{\pm}R^{(f)}_{k}({\bm{p}\over 2}\pm \bm{\ell})}\right.\nonumber\\\fl&&\qquad
\left. -{1-2n_F\left(\xi(\bm{\ell})\right)\over 2\xi(\bm{\ell})}\right]+R^{(b)}_k(\bm{p}). 
\label{eq:4pt_bcs}
\eea
Since it suffices to calculate the coefficient of $p^0$ to determine  $Z_b$, let us put $\bm{p}=0$ in (\ref{eq:4pt_bcs}).
In order to find the properties of low-energy collective excitations, we assume that $k^2/2m$ is much smaller that other energy scales such as the temperature and the Fermi energy. Then, Eq.(\ref{eq:4pt_bcs}) becomes 
\be
{-1\over \Gamma^{(4)}_k(p^0,\bm{0})}=Z_k ip^0 +R^{(b)}_k(\bm{0})+\mathcal{O}(\sqrt{\mu}\cdot k^6/T^3), 
\label{decouple_fermion_4pt}
\ee
where $Z_k$ is given by 
\be
Z_k=\int{\diff^3\bm{\ell}\over (2\pi )^3} {\tanh{\beta\over 2}\left(\xi(\bm{\ell})+R^{(f)}_k(\bm{\ell})\right)\over \left[2\left(\xi(\bm{\ell})+R^{(f)}_k(\bm{\ell})\right)\right]^2} .
\label{eq:wf_bcs}
\ee
Since $Z_k$ converges to a finite value in the limit $k\to0$, we can set $Z_b =Z_0$ in (\ref{eq:boson_reg}) with 
\be
Z_0=\mathrm{p.v.}\int{\diff^3\bm{\ell}\over (2\pi )^3} {\tanh\left({\beta\over 2}\xi(\bm{\ell})\right)\over \left[2\xi(\bm{\ell})\right]^2}.
\label{eq:wf}
\ee
Here, $\mathrm{p.v.}$ denotes the principal-value integral. 
Due to the vertex IR regulator, the collective excitation is also gapped by $k^2/4m$ even when $k^2\ll 2m T$. 
The effect of $R^{(f)}_k$ vanishes of the order of $k^6$ as $k\to 0$. 

In the BEC side where the chemical potential $\mu$ is negative,  $\xi(\bm{q})$ is always positive, so that
 $R^{(f)}_k=0$ for $k<\sqrt{2m|\mu|}\sim 1/a_s$. When $k$ is smaller than $\sqrt{2m|\mu|}$, the inverse propagator of the collective excitation
 is given by 
\bea
&&{-1\over \Gamma^{(4)}_k(p)}=R^{(b)}_k(\bm{p})\nonumber\\
&&-\int{\diff^3\bm{\ell}\over (2\pi )^3} \left[{1-\sum_{\pm}n_F\left(\xi({\bm{p}\over 2}\pm \bm{\ell})\right)\over 2\xi(\bm{\ell}) +(i p^0+{\bm{p}^2\over 4m})} -{1-2n_F\left(\xi(\bm{\ell})\right)\over 2\xi(\bm{\ell})}\right]. 
\eea
Setting $Z_b=Z_0$ with $Z_0$ given in (\ref{eq:wf}), the bosonic excitation becomes gapped by $k^2/4m$. Since the regulator inside the fermion propagator vanishes for small $k$, the vertex IR regulator plays a key role in the BEC side.

For small $k$, the $k$-dependence of $\Gamma_k^{(4)}$ mainly comes from that of the vertex IR regulator so that we find 
\be
\p_k \Gamma_k^{(4)}(p)\simeq \left.\p_{k_2}\Gamma^{(4)}_{k_1,k_2}(p)\right|_{k_1=k_2=k} = \left(\Gamma_k^{(4)}(p)\right)^2\p_k R^{(b)}_k(p).
\label{eq:approx_vertex_flow}
\ee
The neglected term vanishes in the BEC side and is also smaller by a factor $\mathcal{O}(\sqrt{\mu}k^4/Z_0 T^3)$ even for
 positive $\mu>0$ in the BCS side. 
This approximation will play an important role in Sec.\ref{subsec:se_flow} to calculate the momentum dependence of the self-energy. 

\subsection{Self-energy flow and number density}\label{subsec:se_flow}
According to (\ref{flow_se}) and (\ref{vflow_se}), the flow equation for  the self-energy $\Sigma_k(p)=\Sigma_{k,k}(p)$
is given by 
\bea
\p_{k}\Sigma_{k}(p)&=&- \int_l^{(T)} e^{-i l^0 0^+}
\left[{\Gamma_{k}^{(4)}\left(p+l\right) \; \p_k R^{(f)}_k(\bm{l})\over [G^{-1}+R^{(f)}_{k}]^2(l)} 
\right.\nonumber\\
&&\left.\qquad-{\left(\Gamma_{k}^{(4)}\left(p+l\right)\right)^2 \p_{k}R^{(b)}_{k}(\bm{p}+\bm{l})\over [G^{-1}+R^{(f)}_{k}](l)}\right], 
\label{flow_se_nsr}
\eea
where the self-energy corrections in the loop integrals are neglected, while $\Gamma^{(4)}_k$  
 is given in Sec.\ref{subsec:vertex_flow}. 

Let us first analyze the self-energy correction for large  $k$. Since all kinds of excitations are gapped  if $k$ is large, the momentum-dependent part of $\Sigma_k$ must be small. 
In the BCS side, there are gapped fermions inside the Fermi sphere even when $k$ is large, so that the self-energy itself can be comparable with the chemical potential: It only shifts the chemical potential and does not affect the low-energy dynamics. As a result, its effect on $T_c/\ve_F$ is not significant since $T_c$ and $\ve_F$ are shifted in the same way  \cite{Tanizaki:2013doa,Tanizaki:2013fba}. 
In the BEC side, since the number density becomes exponentially small for large $k$, so is the self-energy \cite{Tanizaki:2013yba}.

When $k$ becomes small, we may use the approximation given in (\ref{eq:approx_vertex_flow}), which 
leads to the flow equation of the self-energy (\ref{flow_se_nsr}),
\be
\p_{k}\Sigma_{k}(p)=\int_l^{(T)} e^{-i l^0 0^+}
\left[-{\Gamma_{k}^{(4)}\left(p+l\right) \; \p_k R^{(f)}_k(\bm{l})\over [G^{-1}+R^{(f)}_{k}]^2(l)} 
+{\p_k \Gamma_{k}^{(4)}\left(p+l\right)\over [G^{-1}+R^{(f)}_{k}](l)}\right]. 
\label{flow_se_nsr2}
\ee 
Since the r.h.s. of (\ref{flow_se_nsr2}) is a total derivative in terms of $k$, the solution is found to be
\be
\Sigma_{k}(p)=\int_l^{(T)} e^{-i l^0 0^+}
{ \Gamma_{k}^{(4)}\left(p+l\right)\over [G^{-1}+R^{(f)}_{k}](l)}. 
\label{se_nsr}
\ee

The number density of fermions $n$ can now be evaluated by using the formula (\ref{eq:number}) with the self-energy in (\ref{se_nsr}). 
By taking into account the effect of $\Sigma_0$ up to first-order in (\ref{eq:number}), the number density $n$ is given by 
\bea
n&=&-2\int^{(T)}_p G(p)-2\int^{(T)}_p G(p)^2 \Sigma_0(p) \nonumber\\
&=&-2\int^{(T)}_p G(p)+2\int^{(T)}_{p,l}e^{-i l^0 0^+}G(p)^2 G(l-p) \Gamma^{(4)}_0 (l)\nonumber\\
&=&-2\int^{(T)}_p G(p)+\int^{(T)}_l e^{- i l^0 0^+} {\p \over \p \mu}\ln \Gamma^{(4)}_0 (l). 
\label{number_nsr1}
\eea
To obtain  the last line of  (\ref{number_nsr1}),  we used the following equality given
by taking the $\mu$-derivative of both sides of (\ref{eq:4pt_rpa}),
\be
2\int^{(T)}_l G(l)^2 G(p-l)=\left(\Gamma^{(4)}_0 (p)\right)^{-2}{\p \over \p \mu}\Gamma^{(4)}_0 (p). 
\ee

Eq.(\ref{number_nsr1})
 is nothing but the number equation of the NSR theory, (\ref{eq:nsr_num}), since pp-RPA gives 
\be
{1\over \Gamma_0^{(4)}(p)}={1\over g}+\Pi(p)={m\over 4\pi a_s}+\left(\Pi(p)-{m\Lambda\over 2\pi^2}\right).
\ee 
Thus  we could derive the NSR theory from the MR-FRG flow equations:
The Thouless criterion (\ref{eq:thouless}) and the number equation (\ref{number_nsr1}) give the BCS gap equation (\ref{eq:nsr_gap}) at $T=T_c$ and the number equation (\ref{eq:nsr_num}) of the NSR theory, respectively. 
Accordingly, the equations (\ref{flow_se}-\ref{vflow_4pt}) turn out to be a minimal setup of MR-FRG to describe the entire region of the BCS-BEC crossover. 

\subsection{Number equation from the free energy}\label{app:consistency}
The same expression as (\ref{number_nsr1}) for the number density can be also 
derived from the free energy evaluated in  MR-FRG, 
\be
n={-1\over \beta V}{\p\over \p \mu}\Gamma_0(\beta,\mu), 
\label{eq:app01}
\ee
with $V$ being the volume of the system, and 
 $\Gamma_0(\beta,\mu)/\beta$ being the free energy obtained as the field-independent part of the 1PI effective action $\Gamma_0[\overline{\psi},\psi]$. 

The equation (\ref{form07}) and the flow equation in Fig.\ref{fig:flow_eff2} with the vertex IR regulator give 
\be
\p_k \Gamma_k(\beta,\mu)/\beta V=\int^{(T)}_l {-2 \p_k R^{(f)}_k(\bm{l})\over [G^{-1}-\Sigma_k +R^{(f)}_k](l)}-\int^{(T)}_p \Gamma_k^{(4)}(p)\p_k R^{(b)}_k(\bm{p}). 
\label{eq:app02}
\ee
By taking into account the effect of the self-energy correction in the fermion-loop up to the first order,  the flow equation (\ref{eq:app02}) becomes
\bea
\p_k \Gamma_k(\beta,\mu)/\beta V&=&\int^{(T)}_l{-2 \p_k R^{(f)}_k(\bm{l})\over [G^{-1}+R^{(f)}_k](l)}+\int^{(T)}_l{-2 \Sigma_k(l)\p_k R^{(f)}_k(\bm{l})\over [G^{-1}+R^{(f)}_k]^2(l)}\nonumber\\
\fl&-&\int^{(T)}_p \Gamma_k^{(4)}(p)\p_k R^{(b)}_k(\bm{p}).
\label{eq:app05}
\eea
Substitution of the explicit form (\ref{se_nsr}) into (\ref{eq:app05}) gives 
\bea
&&\p_k \Gamma_k(\beta,\mu)/\beta V=\int^{(T)}_l{-2 \p_k R^{(f)}_k(\bm{l})\over [G^{-1}+R^{(f)}_k](l)}\nonumber\\
\fl&-&\int^{(T)}_p \Gamma_k^{(4)}(p)\left[\int_l^{(T)}{2\p_k R^{(f)}_{k}({\bm{p}\over 2}-\bm{l}) \over \prod_{\pm}\left([G^{-1}+R^{(f)}_{k}]\left({p\over 2}\pm l\right)\right)}+ \p_{k}R^{(b)}_{k}(\bm{p})\right]. 
\label{eq:app06}
\eea
Combining (\ref{flow_vertex_nsr}) and (\ref{eq:app06}), we obtain 
\be
\p_k \Gamma_k(\beta,\mu)/\beta V=-2\int^{(T)}_l{\p_k\ln [G^{-1}(l)+R^{(f)}_k(\bm{l})]}-\int^{(T)}_p \p_k\ln\Gamma^{(4)}_k(p). 
\label{eq:app07}
\ee
Since both sides of (\ref{eq:app07}) are total derivatives in terms of $k$, we obtain the following expression for $\Gamma_k(\beta,\mu)$ up to  a $\mu$ independent constant: 
\bea\fl\hspace{2em}
\Gamma_k(\beta,\mu)/\beta V&=&-2\int^{(T)}_l{\ln [G^{-1}(l)+R^{(f)}_k(\bm{l})]} \nonumber\\&+& \int^{(T)}_p \ln\left[{4\pi a_s\over m}\left(\Gamma^{(4)}_k(p)\right)^{-1}\right]. 
\label{eq:neqF}
\eea
By combining (\ref{eq:app01}) and  (\ref{eq:neqF}), we  obtain the same expression as 
(\ref{number_nsr1}).

\section{Summary and perspectives}\label{sec:summary}
In this paper, we proposed a  new formulation of fermionic FRG  with  multiple infrared regulators, MR-FRG,   and 
  applied the method to the two-component fermionic system which exhibits the BCS-BEC crossover. 
  To control both the fermionic one-particle excitations and the bosonic collective excitations,
  two infrared (IR) regulators were introducded; one for the fermionic two-point function and another for the four-fermion vertex.
   The Nozi\`eres-Schmitt-Rink (NSR) theory, which is successful to capture the 
  qualitative features of the BCS-BEC crossover,  can  be derived  from  MR-FRG.

There are various possibilities to improve the approximations adopted  in MR-FRG to go beyond the NSR theory.
Solving the coupled flow equations for the four-point vertex and the self-energy is necessary  when 
the magnitude of the self-energy becomes comparable with that of the chemical potential.
 The frequency dependence of the four-point function and the higher-point vertex functions would 
 play significant roles in such a situation
 \cite{Tanizaki:2013doa,Tanizaki:2013fba}. (Importance of the frequency dependences is also suggested from the study of unitary Fermi gas in the spin-imbalanced case \cite{PhysRevA.83.063620}.) 
 
  In the BCS regime ($(k_F a_s)^{-1} \ll -1$),  each closed fermion loop for Feynman diagram in the flow equation  produces a factor $n_k\sim k_F^3/3\pi^2$, so that  the number of  closed loops is a good expansion parameter.
This is similar to the hole-line expansion of the Bethe-Brueckner-Goldstone (BBG) theory \cite{Day:1967zza}
for degenerate fermion system: 
 A formal connection between fermionic FRG and the BBG theory  will be reported elsewhere. 
In the BCS regime, the particle-hole loops in the flow of the four-point coupling is  also important to generate
  the Gorkov and Melik-Barkhudarov correction  which
  reduces the critical temperature by a factor $2.2$ from the BCS critical temperature
   \cite{Tanizaki:2013doa,Tanizaki:2013fba, gorkov1961contribution}. 
 The six- or higher-point vertices must be also important to consider the dynamics of collective bosonic excitations in the BCS regime.   
 
   In the BEC regime ($(k_F a_s)^{-1} \gg +1$), the real part of the fermion propagator is always positive and there are no hole excitations.  Nevertheless, the interaction between composite bosons takes place through the fermion closed loops, so that
the critical temperature and the number density are affected through the self-energy and the four-point vertex function. 

Since the present formulation provides a good starting point to make systematic analysis 
of the interacting fermion systems, we expect it to be useful not only for cold atomic systems 
  but also for strongly interacting nuclear/neutron matter and for quark matter.

\section*{Acknowledgements}
The authors  are grateful for useful comments by Gergely Fej\H{o}s. 
T. H. thanks late Gerry Brown for his stimulating discussions on various aspects of  quantum many-body problems in nuclear, hadron and particle physics.    Y. T. was supported by JSPS Research Fellowships for Young Scientists. 
This work was partially supported by RIKEN iTHES project and by the Program for Leading Graduate Schools, MEXT, Japan.
 This work was completed  at the Aspen Center for Physics, which is supported by National Science Foundation grant PHY-1066293.

\bibliographystyle{ws-rv-van}
\bibliography{FRG}

\end{document}